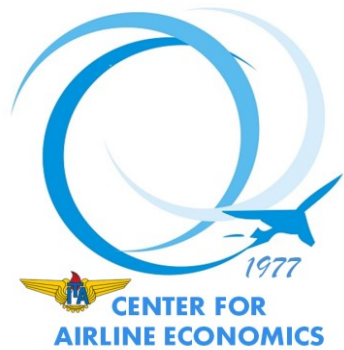

# WORKING PAPER SERIES

Airline delays, congestion internalization
and non-price spillover effects of low cost carrier entry

William E. Bendinelli
Humberto F. A. J. Bettini
Alessandro V. M. Oliveira

# Airline delays, congestion internalization and non-price spillover effects of low cost carrier entry

William E. Bendinelli*

Humberto F. A. J. Bettini†

Alessandro V. M. Oliveira‡

## Abstract

This paper develops an econometric model of flight delays to investigate the influence of competition and dominance on the incentives of carriers to maintain on-time performance. We consider both the route and the airport levels to inspect the local and global effects of competition, with a unifying framework to test the hypotheses of 1. airport congestion internalization and 2. the market competition-quality relationship in a single econometric model. In particular, we examine the impacts of the entry of low cost carriers (LCC) on the flight delays of incumbent full service carriers in the Brazilian airline industry. The main results indicate a highly significant effect of airport congestion self-internalization in parallel with route-level quality competition. Additionally, the potential competition caused by LCC presence provokes a global effect that suggests the existence of non-price spillovers of the LCC entry to non-entered routes.

* Center for Transport Economics, Aeronautics Institute of Technology, Brazil. E-mail: wbendinelli@gmail.com.

† University of São Paulo, Brazil and Center for Transport Economics, Aeronautics Institute of Technology, Brazil. E-mail: humberto.bettini@gmail.com.

‡ Center for Transport Economics, Aeronautics Institute of Technology, Brazil. E-mail: alessandro@ita.br. The authors thank Fapesp, Capes and CNPq for financial support. They also thank Cláudio Jorge P. Alves, Rogéria Eller, Carlos Müller, Márcia Azanha Moraes, Rafael Almeida, Daniel Pamplona, Fernando Capuano, Pâmela Martins, Bruno Arantes, Francisco de Medeiros, Alan Klaubert de Melo, Gianrocco Tucci, Yuri Yegorov, Jose Castillo-Manzano, Anming Zhang and Luke Froeb for useful comments and suggestions. Special thanks to the anonymous reviewers, the Editor and to Jan Brueckner. All mistakes are ours.

**Introduction**

The present paper develops an empirical model to inspect some of the determinants of flight delays and to test their relationships with airline competition and airport dominance. Airline delays have become a constant reality in the modern commercial air travel industry globally. Passengers are now increasingly familiar with flights having light to moderate delays as part of their journey routine. Additionally, episodes of intense flight disruptions due to severe weather conditions, strikes or congestion are observed periodically in many places such as the Eastern and Western United States, Europe, and China, among others. Flight delays and cancellations may not only be stressful to passengers and airlines but are also costly. Moreover, there are costs associated with keeping delays to low rates, and therefore, the motivations to engage in on-time performance also depend on the economic incentives of carriers.

In 2013, in the UK, a congestion charge was under consideration by the government to reduce congestion at both Heathrow and Gatwick to encourage passengers to fly from other London airports like Luton and Stansted[1]. Such an initiative reveals how authorities and operators regard managing congestion and its consequent flight disruptions under a scenario of no airport expansion in the near future. A strand of the airline literature has addressed this issue by inspecting the global effects of airport concentration and dominance on flight delays. Following Daniel (1995), the literature has investigated the *hypothesis of airport congestion internalization*, meaning that a dominant airline could naturally internalize the costs associated with congestion delays that its aircraft impose without the need of a congestion toll – Brueckner (2002), Mayer & Sinai (2003), Daniel & Harback (2008), Rupp (2009), Zhang & Zhang (2006), and Ater (2012).

In addition, in 2014, a report from the Federal Aviation Administration, noted that the absence of competition for many routes might be a source of increased rates of airline flight delays and cancellations[2]. They suggest that competition and service quality may be positively related and therefore more frequent and longer flight delays are likely to be observed on less competitive routes. These findings are in accordance with another strand of the literature on airline delays that investigates the *hypothesis of the competition-quality relationship*. Pioneered by Suzuki (2000), this literature has many recent econometric papers investigating the local, route level determinants of delays, such as Mazzeo (2003), Rupp, Owens & Plumly (2006) and Greenfield (2014). More recently, some articles in this literature have inspected the impacts of the entry of low cost carriers (LCCs) on the on-time performance in the market, as per Rupp (2008), Castillo-Manzano & Lopez-Valpuesta (2014), Bubalo & Gaggero (2015) and Prince & Simon (2015).

---

[1] See "*Passengers at Heathrow and Gatwick could face congestion charge to encourage use of quieter airports*" - The Daily Mail Reporter, 2013, Jan 12.

[2] Office of Inspector General – Federal Aviation Administration (2014). *Reductions in competition increase airline flight delays and cancellations*. Audit Report, n. CR-2014-040, April, 23.

This paper aims to investigate some of the competition-driven incentives of major carriers to keep few records of flight delays, both locally in the market – the *route level* – and globally – the *airport level*. Most papers in the airline delays literature have addressed the subject by focusing on one of these levels in an isolated way. We present a unifying framework to test both the airport congestion internalization hypothesis and the market competition-quality relationship hypothesis in a single econometric model. We examine the role of route and airport concentration metrics as key competition determinants of delays. Another contribution of the paper lies in the modeling of a dynamic pattern of delays after entry by means of a time decomposition into its short-run and long-run effects.

Our primary interest is on the impacts of the entry of LCCs on the odds and average magnitude of flight delays of incumbent FSCs. We investigate the local and global impacts of entry and therefore test the effects of LCC presence on the congestion internalization and local flight service quality of FSCs. We consider the application to the case of the Brazilian airline industry in the period 2002-2013[3]. Our econometric framework addresses the important issue of endogeneity of market structure regressors. Since Greenfield (2014), we have that the magnitude of bias in the estimation of a flight delay equation may be considerable, and therefore, practitioners must implement an instrumental variables approach in such a framework. We consider instrumentation all of the market structure variables and discuss the effect of not accounting for endogeneity.

---

[3] Huse and Oliveira (2012) investigate the impact of Gol's entry on the prices of FSCs, with special focus on the dynamics and the smoothing effect of product differentiation of price responses.

This article is divided as follows: Section 1 presents the conceptual model employed as well as the main hypotheses to be tested; Section 2 presents the empirical model development; Section 3 presents the econometric model and results; Section 4 performs some robustness checks. Finally, the conclusions are presented.

**1. Theoretical framework**

In this section we present our conceptual model along with the main hypotheses investigated in the paper.

*1.1. Airport level determinants of delays: congestion internalization*

The emergence of hub-and-spoke networks is a phenomenon of the post-deregulation period in the US airline industry that has spread to most airline markets in the world. With the formation of hubs, a few carriers a gained dominant position over a set of airports as part of their pro-hub-and-spoke network design strategy. The literature has observed that an airport's dominant airline could have stronger incentives to address congestion than smaller carriers and would therefore naturally internalize the costs associated with its self-imposed flight delays (Daniel, 1995, Brueckner, 2002). This literature focuses on the role of peak/off-peak allocation of flights and passengers at an airport to inspect the incentives to manage congestion and avoid flight delays by dominant carriers. In such situations, congestion tolls aimed at mitigating externalities from flight delays would either not be needed or only be needed in less dominated airports[4]. Recent studies include Mayer & Sinai (2003) and Ater (2012). Based on this strand of the literature, our first hypothesis in the conceptual model is presented below.

**H$_1$. Airport congestion internalization**: airport concentration generates higher incentives for major airlines to engage in congestion internalization.

---

[4] "*One might expect dominant airlines to fully internalize delays of their own aircraft even without congestion pricing.*" – Daniel (1995, p. 333). Ater (2012) explains that dominant airlines may use the length of their flight banks aiming at managing delays: "*With wider banks, flights will interfere less with one another, reducing delays*" (p. 197).

Regarding $H_1$, most of the empirical papers that utilize econometrics find a negative relationship between airport concentration and flight delays. Brueckner (2002) presents rudimentary evidence based on a sample of 25 US airports in 1999; the results are confirmed by Mayer & Sinai (2003) and Ater (2012), who employ panel data disaggregated at the airline-route-time level for the US airline market of the early 2000s. Santos & Robin (2010) present an application to the European airline market from 2000 to 2004 and confirm the findings of congestion internalization. However, Daniel & Harback (2008), Rupp (2009) and, to some extent, Bilotkach and Lakew (2014), find evidence of no self-internalization and therefore suggest a role for congestion pricing in improving economic efficiency. The metrics of concentration utilized in these studies are typically either the airport Herfindahl-Hirschman index (HHI) or the airport share of the dominant firm. Finally, Molnar (2013) finds that internalization depends on the strategic incentives of carriers when balancing the benefits from connections and passenger preferred times with the congestion costs, with evidence that strategic entry deterrence prevails at hubs.

*1.2. Route level determinants of delays: competition and quality*

Another strand of the literature performs a *market level analysis* to inspect the determinants of flight delays. The market level in the airline industry is usually associated with the origin-destination pair – the *route* level. While the airport internalization literature is interested in estimating the *global effects* of concentration – i.e., concerned with the mean airport level – this strand of the literature is interested in estimating the *local effects* of concentration – ie, the market (route) level.

At the route level, some recent papers consider on-time performance as one of the key indicators of airline service quality. Empirical models of delays are specified with metrics of airline competition in the market as regressors, among other factors. Market concentration in this sense is therefore usually viewed as a cause of diminished incentives of carriers to promote service quality. One of the pioneer studies is Suzuki (2000), who finds that on-time performance affects market share on the route through passengers' experience related to delays. The recent econometric literature inverts the analysis and estimates a flight delay equation against market (route) concentration – Mazzeo (2003), Rupp, Owens & Plumly (2006) and Greenfield (2014). All papers find clear evidence that supports a positive competition-service quality relation. Based on the competition-quality strand of the literature, our second hypothesis is therefore as follows.

**H$_2$. Competition-quality relationship**: route (market) concentration generates lower incentives for major airlines to engage in better service quality with respect to on-time performance.

*1.3. A unifying framework: accounting for both the local and global effects of concentration*

A key element of the congestion internalization literature is the focus on the airport level determinants of delays. Indeed, when discussing the role of peak/off-peak allocations on the incentives to internalize congestion, since Brueckner (2002) this literature has been marked by either implicitly or explicitly imposing a *route symmetry assumption* on the empirical modeling. Under route symmetry, all routes out of an airport are virtually equal and therefore a "representative route" can be analyzed without loss of generality[5]. Symmetric routes possess the same market structure and thus local route conditions may easily be extrapolated to the airport level – a useful device considering that route dominance is not sufficient to guarantee airport internalization. Therefore, with symmetry, analyzing either the route or the airport level – the local or the global effects – leads to the same conclusions. Not by coincidence, most papers in this literature utilize airport concentration dominance without recurring to route dominance variables – for example, Brueckner (2002), Mayer & Sinai (2003) and Santos & Robin (2010).

Another issue is related to the competition-quality strand of literature. This literature does not consider the possibility of congestion internalization by airport dominant carriers. Indeed, none of the previous papers utilize airport level concentration metrics to estimate the overall incentives for effective delay management by carriers in their econometric specification – for example, Mazzeo (2003), Rupp, Owens & Plumly (2006) and Greenfield (2014). When restricting it only to the market level analysis, studies consider that competition-driven flight delays are only generated at the local level of the route. Global effects are restricted to nullity, a strong assumption that may be unrealistic if we consider the evidence already produced by the congestion internalization strand of literature.

---

[5] Brueckner (2002, p. 1360) discusses that in his framework, although the number of endpoints is unity, additional endpoints can be added to the model by adding a scale factor by recurring to symmetry across routes.

We believe that an econometric model of flight delays does not need the symmetry assumption of the congestion internalization literature, and thus, we do not impose it in our model. Additionally, we assume the possibility of congestion internalization *in parallel* of service quality management by carriers, which is not assumed by the competition-quality literature. With this framework, we therefore have both market and airport concentration measures in the same empirical equation, each one with its usual interpretations. So far, Mazzeo (2003) and Bubalo & Gaggero (2015) have developed models with both route and airport market structure variables. However, although the studies provide controls for both the local and global dimensions of airline flight delays[6], they do not explicitly discuss the hypothesis of airport congestion internalization in their empirical framework. The apparent disconnection with the congestion internalization research in both papers clearly weakens the possibility of a beneficial interaction of the hypotheses from both the above surveyed strands of literature in their analysis of the results.

### *1.4. The impact of LCCs on internalization and service quality*

In addition to the implementation of a framework in which both route and airport effects of competition may emerge, our primary interest is to estimate the impacts of the entry of low cost carriers (LCCs) on the on-time performance of incumbent full service carriers (FSCs). We investigate the local and global impacts of entry and therefore test the effects of LCC presence on the congestion internalization and flight service quality of FSCs. So far, a few articles in the competition-service quality literature have inspected the impacts of LCCs on flight delays. Rupp (2008) and Castillo-Manzano & Lopez-Valpuesta (2014) find that LCCs have better on-time performance than full-service carriers (FSC). Our framework is similar to Bubalo & Gaggero (2015) and Prince & Simon (2015), which focus on the competitive responses to entry. Based on this recent research, in our framework we propose the following hypothesis and its sub-hypotheses:

---

[6] Mazzeo (2003) use route HHI, airport share and a dummy of city-pair monopoly; Bubalo & Gaggero (2015) use route market share, airport market share – origin, airport market share – destination.

**H₃. LCC entry and responses in flight delays**: The entry of a LCC has impacts on the incentives of incumbent carriers to engage in flight delay reduction.

**H₃ₐ. LCC entry and responses in airport congestion internalization**: LCC entry has impacts on the incentives of incumbent carriers to engage in airport congestion internalization.

**H₃ᵦ. LCC entry and responses in service quality**: LCC entry has impacts on the incentives of carriers to engage in service quality improvement.

Note that differently from **H₁** and **H₂**, we do not impose any restriction on the possible causality relation of LCC entry and flight delays. So far, this literature has been scarce and with conflicting results. Prince and Simon (2015) find evidence that LCC entry increases flight delays of incumbent airlines, as it forces a vigorous price competition that induces carriers to cut costs – one of such costs being the costs associated with on-time performance management. Bubalo & Gaggero (2015) find evidence of the contrary, however. We therefore believe that the impacts of competitive pressure stemming from LCCs on delays lack consensus and are still an empirical matter to be further investigated. Additionally, the airport congestion internalization literature so far has completely neglected the LCC entry issue.

The role of our hypotheses above may be better visualized in Figure 1, which presents our conceptual model, with the representation of the key drivers of airline flight delays and their interactions. As shown in Figure 1, we regard flight delays as being directly incurred by airlines but acknowledge the fact that delay is produced by a combination of interactions of air transport players – namely, the airline, the airport and the air traffic control (ATC) entities. Additionally, external factors such as bad weather, strikes and incidents/accidents must be accounted for in any analysis of airline on-time performance. Our focus here is on the airport-airline interactions through vertical relationships – which may cause congestion internalization – and on the market-airline (route) interaction – which may produce service quality competition. Our two first hypotheses, **H₁** and **H₂**, are designed to explicitly model these local and global effects of competition on flight delays. Our third hypothesis (**H₃**), regarding LCC entry, enters the model through both a market-airport-airline relationship and a market-airline relationship.

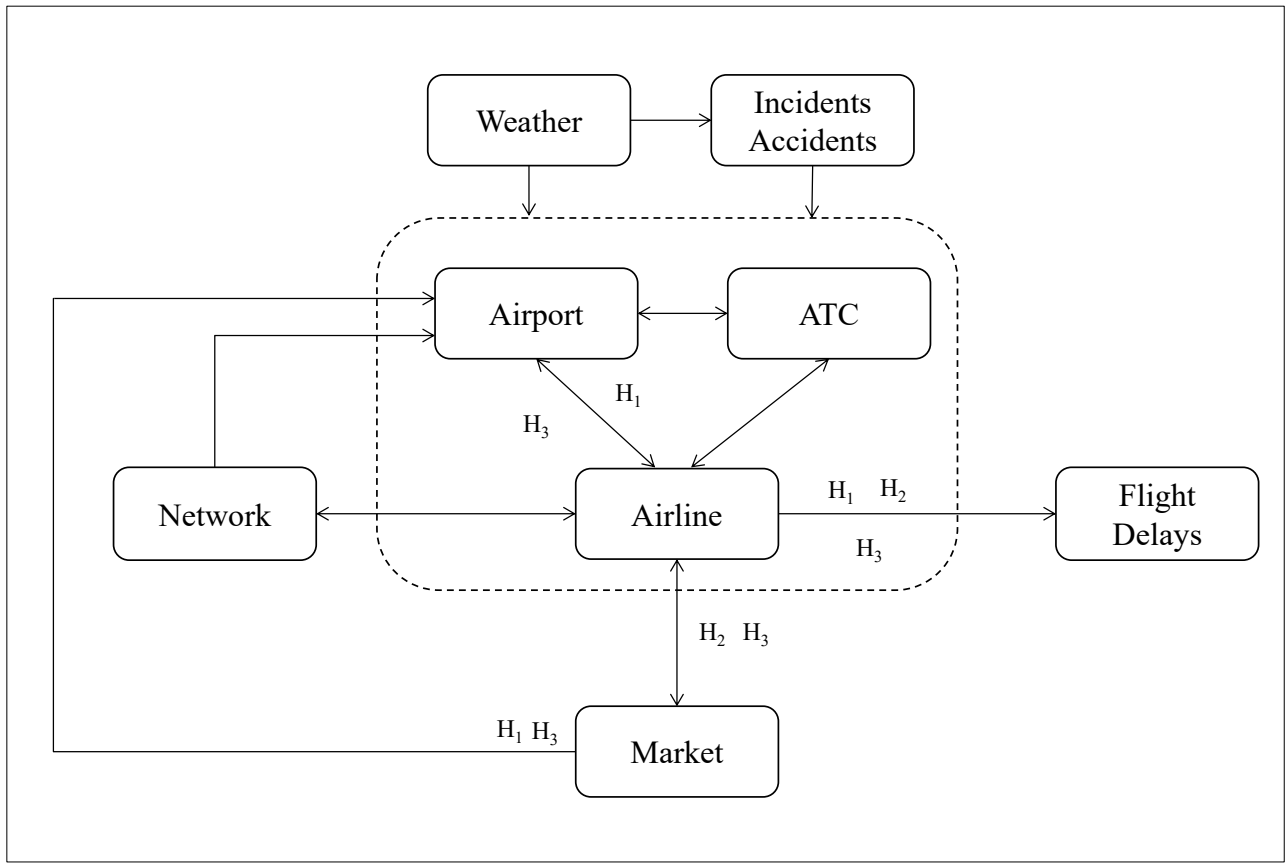

**Figure 1 – Conceptual model of airline delays**

## 2. Empirical model development

### *2.1. Application*

We develop an empirical model of flight delays for the Brazilian airline industry. Brazil constitutes a case in which airline delays have long been intensively discussed and for which congestion tolls have not been implemented so far. In 2008 three of the most important airports in the country – Brasília International Airport (BSB), São Paulo/Congonhas (CGH) and São Paulo/Guarulhos (GRU) – were among the five most delayed airports in the world[7]. The country was forced to engage in a major effort of flight operations oversight along with a regulatory reform regarding punishing long flight delays and managing scarce airport slots to avoid problems during the 2014 FIFA World Cup[8]. Years before, commercial aviation in Brazil was fully deregulated in 2001, with the institutional arrangement for airport slot allocation first introduced in 2006. So far, only São Paulo's downtown Airport of Congonhas (CGH) has been officially designated as a coordinated airport, with strict slot allocation rules dictated by the independent authority National Agency of Civil Aviation (ANAC). The allocation mechanism is marked by a grandfather rights system of "use-it or lose-it" but with some recent pro-new entrants regulation[9]. The remaining airports are not subject to slot allocation rules. The International Airport of São Paulo/Guarulhos (GRU) is currently a schedules facilitated airport and participates in the IATA Worldwide Scheduling Guidelines and Conference.

---

[7] "*The World's most-delayed airports*" (Forbes, 2008, Jan 14).

Since 2001, a number of structural changes have been observed in the industry, examples being the birth of the low-cost carriers Gol Airlines in 2001 and Azul Airlines in 2008, the rise and fall of a major strategic alliance – the codeshare agreement of Varig and Tam airlines in 2003-2005 - and more recently, the privatization of key airports since 2012. The major changes in the airline market through the 2000s have produced positive and negative aspects. Between 2002 and 2010, according to the National Civil Aviation Agency, domestic air transport presented a 153% increase in revenue-passenger kilometers and a 52% drop in average yields. In parallel, the airlines have notably designed marketing strategies to attract consumers such as the emerging new middle class. However, the rapid growth in demand was concomitant with the concentration on a few major hubs such as the São Paulo city airports, thus causing pressure on the existing airport infrastructure.

After the bankruptcy of flag carrier Varig, the market structure became rapidly concentrated. Market deconcentration had restarted only in 2008, with the entry of LCC Azul Airlines, with its smaller-than-average aircraft types through the use of Embraer's E-Jets, along with the intense utilization of the underexplored niche of secondary airport operations out of São Paulo/Campinas Airport (VCP). Airport congestion caused by infrastructure bottlenecks across the country was evident throughout the 2000s and, in particular, in the "air blackout" period of 2006-2007 in which more than one-third of the flights were disrupted, and later with the rapid acceleration in economic growth from 2010. The lack of airport competition and the meager public budget for improvements and expansions meant that airport infrastructure scarcity has still been an issue in the country. Since the blackout period, however, flight delays and cancellation have not been a recurrent problem for authorities, as the proportion of flight disruptions declined considerably. Table 1 permits observing the evolution of flight delays and their possible association with airline hubbing (connecting passengers) and concentration both at the market (route) and the city (mean airport) level.

---

[8] "*World cup Brazil flight delays to result in airline fines*" (Bloomberg Business, 2014, Jan 23).

[9] For example, in October 2014, 100% of the new slots available at the airport on the occasion were distributed to new entrants. See "*Brazil ANAC announces slot changes at São Paulo Congonhas Airport, increase from 30 to 32-33 per hr*" (CAPA – Center for Aviation, 29-Sep-2014).

From Table 1 we know that the proportion of delayed flights decreased considerably if we compare the beginning of the 2010s with any period of the 2000s. When contrasted with the second half of the previous decade, which consists the 2006-2007 air blackout period, flight delays decreased by one-third (33.5%). When compared with a less abnormal period such as the 2002-2005 years (the bottom line of Table 1), the decrease in the proportion of delayed flights was 5.3% on average. Additionally, considering the same comparison it is possible to observe an 11% reduction in hubbing – measured by the proportion of connecting passengers – and a 4.3% increase in city HHI. This analysis suggests that some airport congestion internalization may have occurred. In contrast, market (city-pair) HHI decreased by 0.4%, and thus suggesting a slight competition-quality positive relationship.

**Table 1 – Flight delays (arrivals), hubbing (connections), and market concentration in Brazil (2002-2013)[10]**

| Period | Delayed flights (%) | Connecting passengers (%) | Herfindhal-Hirschman (HHI) | |
|---|---|---|---|---|
| | | | City-Pair | City |
| (1) 2002-2005 | 21.1% | 12.5% | 0.473 | 0.332 |
| (2) 2006-2010 | 30.0% | 11.3% | 0.527 | 0.404 |
| (3) 2011-2013 | 19.9% | 11.1% | 0.471 | 0.346 |
| % Var. (3)/(2) | -33.5% | -1.9% | -10.6% | -14.3% |
| % Var. (3)/(1) | -5.3% | -11.0% | -0.4% | 4.3% |

---

[10] Source: National Civil Aviation Agency, VRA report, 2002-2013; Infraero, unpublished monthly airport movement report, 2002-2013; own calculations. Delayed flights are accounted when a flight is delayed by more than fifteen minutes. HHI figures measured using market share of revenue passengers.

*2.2. Data*

Our dataset consists of a panel of 209 routes in Brazil between January 2002 and December 2013. The dataset comprises only routes involving the Brazilian state capitals and the country's capital[11]. In our analysis, a route is defined as a domestic directional city-pair[12]. Most data utilized in this research are publicly available from the National Civil Aviation Agency (ANAC). ANAC is responsible for monitoring punctuality and regularity of all domestic flights operated by scheduled airlines in Brazil. Detailed information of all scheduled flights in the country is available in an online database named Active Scheduled Flight Report (VRA). VRA contains flight level data of carrier, airport-pair, flight number, and scheduled and actual departure and landing times since 2000. The database also presents the justification code reported for each delayed and cancelled flight – bad weather conditions, incidents, etc. A limitation of the data set is related to the cases of multiple causes of delay. When a given flight is delayed or cancelled due to multiple causes, only the most relevant reason is reported. These data are provided by airlines and, on account of the potential strategic incentives in reporting reasons for delays, it is subject to inspection and audit by ANAC. For the Brazilian authority, a given flight is considered delayed when it arrived 30 or more minutes than the schedule. For our purposes in this paper, we use the 15 minutes standard of the U.S. Department of Transportation's (DOT) Bureau of Transportation Statistics (BTS), also used by most of the literature.

The original dataset contains information of 10 million flights collected from ANAC's VRA report. We restrict our attention to the subset of the scheduled full-service carriers (FSCs) of the database, namely Tam, Varig, Transbrasil and Vasp. Gol and Azul airlines are the low cost carriers (LCCs) in the sample. Additionally, we aggregate the dataset to the route-month level to inspect the determinants of average flight delays of FSCs. Socio-economic data were collected from the Brazilian Institute of Geography and Statistics (IBGE) and the Brazilian Central Bank.

---

[11] There are 26 state capitals. We also included the country's capital, Brasília. In this set of cities, the minimum observed population is 230 thousand people.

[12] In our city-pair setting, the airport of Guarulhos International (GRU) and Campinas/Viracopos International (VCP) are considered as belonging to São Paulo multiple airports region. Additionally, Confins International (CFN) belongs to Belo Horizonte city.

## 2.3. Empirical model

Equation (1) presents our empirical model of flight delays in the Brazilian airline industry:

$$\begin{aligned}
ODDS_{kt} = &\ \beta_1 nr\ flights\ in\ congested\ hours_{kt} \\
&+ \beta_2 nr\ flights\ in\ uncongested\ hours_{kt} \\
&+ \beta_3 prop\ flights\ with\ bad\ weather_{kt} + \beta_4 prop\ flights\ with\ incidents_{kt} \\
&+ \beta_5 prop\ flights\ held\ for\ late\ connections_{kt} \quad (1) \\
&+ \beta_6 max\ prop\ city\ delayed\ flights_{kt} + \beta_7 codeshare\ agreement_{kt} \\
&+ \beta_8 HHI\ city\text{-}pair_{kt} + \beta_9 HHI\ max\ endpoint\ cities_{kt} \\
&+ \beta_{10} LCC\ pres\ city\ pair_{kt} + \beta_{11} LCC\ pres\ max\ endpoint\ cities_{kt} \\
&+ \gamma_k + \gamma_t + u_{kt}
\end{aligned}$$

Where:

- $ODDS_{kt} = \ln[\text{prop delayed flights}_{kt}/(1 - \text{prop delayed flights}_{kt})]$, where prop delayed flights$_{kt}$ is the proportion of FSC flights reported with delays on city-pair $k$ and time $t$. Source: National Civil Aviation Agency, VRA Report, with own calculations. A delay is computed whenever the difference between the scheduled arrival time and the actual arrival time is higher than fifteen minutes. Alternative specifications considering departure delays were also utilized to check the robustness of the results.

- $nr\ flights\ in\ congested\ hours_{kt}$ is the average number of daily scheduled flights of city-pair $k$ and time $t$ that operate during congested hours. A "congested hour" is defined as a full clock hour characterized by operations of flights (arrivals plus departures) in a higher amount than official declared capacity. Sources: National Civil

Aviation Agency, VRA Report and an airport capacity study commissioned by the Brazilian government (2010)[13].

- $nr\ flights\ in\ uncongested\ hours_{kt}$ is the average number of daily scheduled flights of city-pair $k$ and time $t$ that operate in uncongested hours – see the definition of "congested hour" above. Source: National Civil Aviation Agency, VRA Report and Brazilian government.

- $prop\ flights\ with\ bad\ weather_{kt}$ is the proportion of flights delayed with the justification of operations under bad weather conditions of city-pair $k$ and time $t$. This variable has the total number of actual flights on a city-pair as the denominator. Source: National Civil Aviation Agency, VRA Report.

- $prop\ flights\ with\ incidents_{kt}$ is the proportion of flights delayed with the justification of operational incidents of city-pair $k$ and time $t$. This variable has the total number of actual flights on a city-pair as the denominator. Source: National Civil Aviation Agency, VRA Report.

- $prop\ flights\ held\ for\ late\ connections_{kt}$ is the proportion of flights delayed with the justification of awaiting passengers or load from another flight of city-pair $k$ and time $t$. This variable has the total number of actual flights on a city-pair as the denominator. Source: National Civil Aviation Agency, VRA Report.

- $max\ prop\ city\ delayed\ flights_{kt}$ is the maximum proportion of delayed flights between the origin and destination endpoint cities of city-pair $k$ and time $t$. This variable has the total number of actual flights on a city as the denominator. This variable is designed to capture the overall vulnerability of endpoint cities to shocks in the whole network. In particular, with this variable, we aim to control for unobserved delay propagation effects that may be caused by exogenous shocks and also account for overall system wide conditions, such as during the blackout period. Source: National Civil Aviation Agency, VRA Report.

---

[13] "*Study of the Air Transport Sector in Brazil*" (text in Portuguese) - Brazilian Development Bank, Jan, 25, 2010, available at www.bndes.gov.br.

- *codeshare agreement*$_{kt}$ is a dummy variable to account for the city-pairs and periods in which the codeshare agreement between the major carriers TAM and Varig had operations. The 2003-2005 codeshare agreement was the only relevant alliance among airlines in the sample period. Source: Secretariat for Economic Monitoring (SEAE) of the Ministry of Finance.

- *HHI city-pair*$_{kt}$ is the Herfindahl-Hirschman index of concentration of revenue passengers of city-pair $k$ and time $t$. To extract this variable, we use the city-pair level market shares of all participating carriers and then calculate the city-pair level concentration. This variable aims at capturing the effect on delays of airline market dominance, ie. market concentration at the city-pair level. Source: National Civil Aviation Agency, Monthly Traffic Report, with own calculations.

- *HHI max endpoint cities*$_{kt}$ is the maximum Herfindahl-Hirschman index of concentration of revenue passengers between endpoint cities of city-pair $k$ and time $t$. In other words, it is a proxy for the concentration measured at the city level. To obtain this variable, we use the city-level market shares of all participating carriers to extract the city-level concentration. For each city-pair market, we compute the maximum city-level concentration between the respective origin and destination cities. This variable aims at capturing the effect on delays of overall dominance of the available airports of a city. We utilized the maximum city concentration between origin and destination as in Boguslaski, Ito and Lee (2004). The justification for using the maximum value lies in the fact that it is typically enough to have one of the endpoints cities concentrated to start producing effects of dominance over the market[14]. We also experimented with the passenger-weighted average HHI at the two endpoint cities, with the final results being robust to this change. Source: National Civil Aviation Agency, Monthly Traffic Report, with own calculations.

- *LCC presence city pair*$_{kt}$ is a dummy variable to account for the presence of LCCs Gol and Azul airlines on city-pair $k$ and time $t$. Source: National Civil Aviation Agency, Traffic Report.

---

[14] Note that our dataset does not contain small cities that are also typically concentrated due to natural monopoly characteristics.

- *LCC presence max endpoint cities*$_{kt}$ is a dummy variable to account for the presence of LCCs Gol and Azul airlines on either origin or destination endpoint cities of city-pair $k$ and time $t$. The LCC city presence measure is conceived to work as the usual potential competition measure, i.e., presence at an endpoint but not on the route itself still provides competitive pressure. Using the maximum operator allows us to consider any route presence at the airports of a city, besides being consistent with our max city HHI notation. Source: National Civil Aviation Agency, Traffic Report.

- $\beta_1, \ldots, \beta_{11}$ are unknown parameters.

- $\gamma_k$ and $\gamma_t$ are, respectively, the city-pair and time fixed effects.

- $u_{kt}$ is the disturbances term, of which we provide more details in 2.4.

In an alternative specification of (1), we also consider the following regressand:

- $MINS_{kt}$, the mean difference in minutes between scheduled arrival time and actual arrival time of FSCs on city-pair $k$ and time $t$ - a metric that may be negative. We also consider truncated versions of $MINS_{kt}$, considering only positive values or figures above a certain threshold delay (for example, $MINS_{kt} > 15$ meaning delays above fifteen minutes; we also experiment with thirty minutes). We also consider the terminology of Greenfield (2014) – $ODDSD_{kt}$ and $MINSD_{kt}$ to measure departure delays in a set of robustness checks.

One issue to emphasize is that our flight delays equation has variables measured both at the city-pair level and the city level. This is consistent with the fact that a flight on a particular route most likely will not ending being congested by another flight on that *same* route but, instead, it will most likely be stuck behind a flight on a *different* route in waiting to take off or landing. As all routes from/to a given city share the same terminal control area, it is likely that flight delays will be generated not only by the specificities of a route but the general conditions of all city airports. This observation aims at making the distinction clearer between delays due to route characteristics and delays affected by overall city characteristics.

Note that our concentration measures *HHI city-pair*$_{kt}$ and *HHI max endpoint cities*$_{kt}$ are extracted consistently with our definition of market as being the city-pair level instead of the airport-pair level. Brueckner, Lee and Singer (2014), provide evidence that city-pairs, instead of airport-pair, are the appropriate market definition in many cases of air transportation analyses. By restricting our attention to the city-pair level, however, we are aware that this setup may compromise some of the ability to understand interactions among routes at the airport-level that could be important in airline response to congestion. In particular, we believe that competition in flight frequencies may produce the result of carriers strategically moving flights from adjacent airport-pairs to strengthen market position at a given airport pair. With our setup, we are not able to observe such strategic movements that may be related to market concentration and congestion. We stress that, from the 27 cities present in the database, only three are in fact multiple airport regions for scheduled flights: São Paulo, Rio de Janeiro and Belo Horizonte. In all cases, the city airports belong to the same terminal control area and therefore are subject to the same congestion avoiding approached for the controlled airspace. Additionally, because all major carriers in Brazil are usually present in all markets, the concentration level measured at the city-pair (or average city) level and at the airport-pair (or average airport) level are typically highly correlated. Due to the resulting strong multicollinearity, it is not possible to estimate a *ceteris paribus* effect for variables at both airport and city levels. Finally, we interpret our setup with city-pair and city level variables as a way to at least partially control for both airport-pair and airport level phenomenon. For example, if market concentration increases in a given airport-pair by 10%, we expect the market concentration of the city-pair also to increase unless the adjacent airport-pairs have concomitant decreases in market concentration. With a 10% increase in airport-pair concentration and holding adjacent pairs constant, the overall city-pair concentration would obviously increase by less than 10%. Consequently, variables measured at the city-pair level will typically contain less sample variability and therefore are more difficult to present statistically significant results. We consider this characteristic conservative – and desirable in a sense – from the econometric standpoint[15].

A final caveat of our analysis is related to the measurement of flight delays. We are aware of the limitations of our procedure of measuring delays strictly relative to flight schedules. Particularly in the situation of congested airports and congested periods, carriers may engage in strategic movements related to the inclusion of padding of schedules[16]. Actually, under schedule padding, the imbedded buffers may implicitly include both congestion-related departure and arrival delays, ultimately resulting in "on time" flights in our dataset[17]. The way the literature so far has addressed the issue of schedule padding follows the procedure of Mayer and Sinai (2003). Instead of calculating delays based on the difference between actual and scheduled arrival and departure times, the authors utilize the "excess travel time" - the difference between actual travel time and the minimum travel time by the carrier on the route. This procedure is not immune to criticism, however. Rupp (2009), for example, argues that it is highly unlikely that passengers calculate excess travel time in their perceptions of the length of flight delays. Here we follow the more traditional approach to delays but recognize that accounting for the strategic schedule padding of airlines may be a necessary extension to our model.

---

[15] We recommend that future research applied to cases with a higher number of multiple airport areas to double check the results of the empirical model utilizing a city-pairs database with the results obtained from a database disaggregated at the airport-pair level.

From now, we omit indexes $k$ and $t$. Table 2 presents descriptive statistics of the sample.

**Table 2 - Descriptive statistics**

| Variable | | (1) | (2) | (3) | (4) | (5) | (6) | (7) | (8) | (9) | (10) | (11) | (12) | (13) |
|---|---|---|---|---|---|---|---|---|---|---|---|---|---|---|
| **Pearson Correlation** | | | | | | | | | | | | | | |
| nr flights in congested hours | (1) | 1.00 | | | | | | | | | | | | |
| nr flights in uncongested hours | (2) | 0.45 | 1.00 | | | | | | | | | | | |
| prop flights with bad weather | (3) | -0.03 | -0.03 | 1.00 | | | | | | | | | | |
| prop flights with incidents | (4) | -0.04 | -0.03 | 0.02 | 1.00 | | | | | | | | | |
| prop flights held for late connections | (5) | -0.04 | -0.03 | -0.03 | 0.00 | 1.00 | | | | | | | | |
| max prop city delayed flights | (6) | -0.05 | -0.09 | 0.61 | 0.08 | 0.09 | 1.00 | | | | | | | |
| codeshare agreement | (7) | -0.01 | -0.09 | -0.22 | -0.03 | 0.03 | -0.25 | 1.00 | | | | | | |
| HHI city-pair | (8) | -0.21 | -0.32 | 0.11 | -0.01 | -0.04 | 0.18 | 0.02 | 1.00 | | | | | |
| HHI max endpoint cities | (9) | -0.13 | -0.13 | 0.21 | -0.03 | 0.07 | 0.32 | -0.23 | 0.48 | 1.00 | | | | |
| LCC presence city-pair | (10) | 0.09 | 0.16 | 0.11 | -0.05 | 0.10 | 0.07 | -0.25 | -0.40 | -0.08 | 1.00 | | | |
| LCC presence max endpoint cities | (11) | 0.01 | 0.02 | 0.00 | -0.05 | 0.02 | 0.02 | -0.08 | -0.03 | -0.07 | 0.11 | 1.00 | | |
| ODDS | (12) | 0.01 | -0.02 | 0.72 | 0.15 | 0.11 | 0.58 | -0.12 | 0.03 | 0.15 | 0.05 | -0.03 | 1.00 | |
| MINS | (13) | -0.02 | -0.12 | 0.57 | 0.12 | 0.03 | 0.49 | 0.08 | 0.16 | 0.09 | -0.12 | -0.04 | 0.70 | 1.00 |
| **Univariate statistics** | | | | | | | | | | | | | | |
| Mean | | 1.68 | 7.87 | 0.18 | 0.01 | 0.03 | 0.24 | 0.18 | 0.48 | 0.40 | 0.90 | 1.00 | -1.38 | 7.16 |
| Standard Deviation | | 5.50 | 11.29 | 0.13 | 0.02 | 0.04 | 0.10 | 0.39 | 0.15 | 0.07 | 0.30 | 0.04 | 1.03 | 8.29 |
| Minimum | | 0.00 | 0.0 | 0.00 | 0.00 | 0.00 | 0.05 | 0.00 | 0.21 | 0.23 | 0.00 | 0.00 | -4.90 | -9.80 |
| Maximum | | 78.84 | 115.8 | 0.96 | 0.3 | 0.65 | 0.70 | 1.00 | 1.00 | 1.00 | 1.00 | 1.00 | 4.03 | 131.91 |

---

[16] See "*Why a six-hour flight now takes seven*" (by Scott Mccartney, The Wall Street Journal, Feb. 4, 2010).

[17] See Villemeur et al (2014) for a recent work on the issue.

## 2.4. Estimation strategy

We performed tests of heteroscedasticity and autocorrelation in the residuals. Firstly, the Pagan-Hall, White/Koenker and Breusch-Pagan/Godfrey/Cook-Weisberg heteroscedasticity tests, with alternative specifications of levels, squares, cross products of regressors and also fitted values of the regressand. All these tests strongly rejected the null of homoscedastic disturbances. We also implemented a Cumby-Huizinga test of autocorrelation for several order specifications, already accounting for heteroscedasticity and endogeneity[18]). These tests indicated the presence of autocorrelation of order 13. We employed the procedure of Newey-West to adjust the standard error estimates[19].

With respect to the endogeneity of some of the regressors we know that, consistent with our conceptual model, namely, the airline-airport-market relationship, the market structure variables are endogenously determined and therefore correlated with the error term. This procedure is consistent with Greenfield (2014). We therefore treat the endogeneity of both HHI and max city HHI by employing an instrumental variables estimator, the Two-Step Feasible Efficient Generalized Method of Moments (2SGMM) with standard errors robust and efficient to arbitrary heteroscedasticity and autocorrelation. The setup of the estimator employed a Newey-West (Bartlett) kernel and a fixed-effects procedure with seasonality controls as discussed before. We also present a study of the problems of disregarding the endogeneity issue, i.e. utilizing Ordinary Least Squares (OLS).

Our identification strategy considers Hausman-type instruments[20] as in Piga & Bachis (2006) and Mumbower, Garrow & Higgins (2014). With Hausman-type instruments, we utilize concentration levels of other routes to instrument the concentration level of a given route. The identifying assumption of the Hausman-type instruments permits exploiting the panel structure of the data by assuming that concentration levels are correlated across markets but independent of each other's unobserved shocks. We suspect that common shocks may be a reality, however, for routes out of endpoint cities more geographically related. The more distant the endpoint cities of two given routes the more realistic would be the use of

---

[18] On the issue of endogeneity, see the discussion below.

[19] As discussed by Baum, Schaffer and Stillman (2007), we utilized the Bartlett kernel function with a bandwidth of $T^{1/3}$, where $T = 144$.

[20] See Hausman (1996).

Hausman-type instruments. Considering this reasoning, we therefore discard nearby cities when computing the mean concentration levels of all other routes to instrument the concentration level of a given city. We employ three cut-off thresholds, 150, 300 and 500 kilometers, to produce alternative instruments and challenge the validity and relevance of them with statistical tests. First, the validity of the full set of over identifying conditions was analyzed by utilizing Hansen J tests. Rejection of the null hypothesis implies that instruments are not satisfying the orthogonality conditions, one reason being that they are not truly exogenous. For most considered specifications, the Hansen J tests did not reject orthogonality. Second, the relevance of the proposed set of instruments was challenged by underidentification tests. The test employs the Kleibergen-Paap rk LM statistic (KP). The tests led to the rejection of the null of underidentification. Finally, we also tested for weak identification. Considering both the Cragg-Donald Wald F statistic and the Kleibergen-Paap rk Wald F statistic (Weak CD and Weak KP), we had evidence for rejecting the hypothesis of weak instruments. The results of all performed tests on the quality of instruments are reported in the tables of Section 3.

## 3. Results

Table 3 presents the estimation results of our empirical model of flight delays in Brazil. We consider three different specifications of the regressand: $ODDS$, $MINS$, and $MINS > 15$ minutes. Columns (1), (3) and (5) contain our baseline models[21].

Some important findings may be obtained from Table 3. First, in all models we have clear evidence of airport congestion internalization - our first hypothesis, $H_1$. Indeed, the results for the estimated coefficients of $HHI\ max\ endpoint\ cities$ are negative and statistically significant in all $ODDS$ and $MINS$ specifications. That result confirms the findings of Brueckner (2002), Mayer & Sinai (2003), Santos & Robin (2010), and Ater (2012). Second, with respect to the competition-quality hypothesis ($H_2$), we also find evidence confirming the expected theoretical relationship between delays and route concentration of Mazzeo (2003) and Greenfield (2014). Indeed, in all cases the coefficients of $HHI\ city\text{-}pair_{kt}$ are positive and statistically significant at least at the 5% level. Third, regarding the impacts of LCC entry – sub-hypotheses $\mathbf{H_{3a}}$ and $\mathbf{H_{3b}}$ – we find some evidence that incumbent FSCs engage in extra internalization following entry with respect to the odds of flight delays, but not to the length of such delays, as the negative coefficient of $max\ city\ LCC\ presence$ is statistically significant in the $ODDS$ equation but not in the $MINS$ equations. Local (route) level responses to entry, inspected with $LCC\ presence$, are either not significant or only significant at the 10% level. We therefore do not find enough evidence supporting the cost/price cutting hypothesis of Prince and Simon (2015).

---

[21]As suggested by one anonymous reviewer, we experimented inserting a variable calculated as the absolute difference between the average airport-pair HHI and the city-pair HHI, in order to inspect whether there is additional competition through close substitutes. This variable was not statistically significant in most cases and results were not changed. We believe that our LCC variables already capture this important effect because the operational base of the LCC Azul Airlines is actually São Paulo/Viracopos - the only relevant secondary airport in the country.

Our results suggest the conclusion that self-internalization of airport congestion is observed in the Brazilian airline market and is also induced by LCC entry. Following entry, the city share of dominant carriers decreases and the consequent drop in concentration tends to provoke a reduction in congestion internalization and an increase in flight delays. In this sense, our estimates show that the presence of the LCC in a city works as a moderating effect on such reduction of congestion internalization. A depeaking strategy may be used as an attempt to keep internalizing congestion even with decreasing airport concentration, which may be accomplished by reducing the complexity of the hub and spoke operations when competing with the LCC. Flights may be allocated in off-peak hours in which the LCC is more attractive to leisure passengers – weekends, for example. Other alternative strategies aimed at better on-time performance are also possible to justify our results. For example, unobserved improvements in network and scheduling management of the incumbent after LCC entry may induce permanent decreases in flight delays. These results are consistent with an overall depeaking movement by FSCs that reduces overall congestion. Entry on a route therefore generates potential competition on the other routes of a city, and thus results in a positive spillover effect that benefits non-entered routes toward better on-time performance. Note, however, that we did not find enough evidence giving support to either a moderating effect or an accentuation effect of the LCC entry on the competition-quality relation. We therefore did not observe any *ceteris paribus* effect of the LCC entry on flight delays at the route level, apart from the global effect of extra internalization caused by entry on the prevalence of delays.

**Table 3 – Estimation results**[22]

|  | (1) ODDS | (2) ODDS | (3) MINS | (4) MINS | (5) MINS > 15 | (6) MINS > 15 |
|---|---|---|---|---|---|---|
| nr flights in congested hours | 0.0043** | 0.0043* | 0.0801* | 0.0827* | 0.0778* | 0.0810* |
|  | [0.002] | [0.002] | [0.041] | [0.043] | [0.041] | [0.043] |
| nr flights in uncongested hours | 0.0046** | 0.0044** | 0.0450 | 0.0571 | 0.0451 | 0.0589 |
|  | [0.002] | [0.002] | [0.028] | [0.036] | [0.028] | [0.036] |
| prop flights with bad weather | 4.7119*** | 4.7145*** | 31.7681*** | 31.6405*** | 31.1137*** | 30.9731*** |
|  | [0.092] | [0.093] | [1.247] | [1.258] | [1.254] | [1.265] |
| prop flights with incidents | 6.2157*** | 6.1807*** | 50.0528*** | 50.5325*** | 48.4526*** | 48.9637*** |
|  | [0.394] | [0.395] | [4.998] | [5.005] | [5.112] | [5.115] |
| prop flights held for late connections | 2.4272*** | 2.4781*** | 22.0148*** | 19.5129*** | 21.9546*** | 19.3264*** |
|  | [0.275] | [0.257] | [5.062] | [4.440] | [5.125] | [4.490] |
| max prop city delayed flights | 1.5564*** | 1.5460*** | 12.5036*** | 12.6235*** | 12.3394*** | 12.4196*** |
|  | [0.225] | [0.228] | [2.893] | [2.894] | [2.928] | [2.933] |
| codeshare agreement | 0.0207 | 0.0081 | 1.1053 | 1.3287 | 1.0843 | 1.3123 |
|  | [0.061] | [0.061] | [0.897] | [0.866] | [0.901] | [0.870] |
| HHI city-pair | 0.8050** | 0.8192** | 30.6290*** | 30.0753*** | 31.9607*** | 31.5567*** |
|  | [0.373] | [0.410] | [8.617] | [8.853] | [8.778] | [9.041] |
| HHI max endpoint cities | -1.4772*** | -1.5144*** | -19.4278*** | -19.1045*** | -20.9849*** | -20.6986*** |
|  | [0.523] | [0.527] | [6.583] | [6.514] | [6.684] | [6.615] |
| LCC presence city-pair |  | -0.0412 |  | 2.4889* |  | 2.7123* |
|  |  | [0.072] |  | [1.464] |  | [1.497] |
| LCC presence max endpoint cities |  | -0.4234** |  | -0.4861 |  | -0.8230 |
|  |  | [0.179] |  | [1.683] |  | [1.706] |
| city-pair fixed effects | yes | yes | yes | yes | yes | yes |
| time fixed effects | yes | yes | yes | yes | yes | yes |
| Adj. R-Squared | 0.6801 | 0.6800 | 0.4546 | 0.4629 | 0.4374 | 0.4456 |
| RMSE Statistic | 0.5889 | 0.5890 | 6.1713 | 6.1244 | 6.2268 | 6.1813 |
| F Statistic | 79.005 | 76.576 | 26.525 | 26.879 | 25.221 | 25.561 |
| KP Statistic | 154.2698 | 139.2305 | 30.7876 | 29.8792 | 30.7876 | 29.8792 |
| KP P-Value | 0.0001 | 0.0001 | 0.0001 | 0.0001 | 0.0001 | 0.0001 |
| J Statistic | 3.1132 | 3.2199 | 0.2161 | 0.6514 | 0.1447 | 0.5509 |
| J P-Value | 0.3745 | 0.3589 | 0.6420 | 0.4196 | 0.7036 | 0.4579 |
| Weak CD Statistic | 91.0127 | 83.4722 | 35.3711 | 37.2128 | 35.3711 | 37.2128 |
| Weak KP Statistic | 34.0972 | 30.5968 | 10.3854 | 10.0724 | 10.3854 | 10.0724 |
| Nr Observations | 19419 | 19419 | 19590 | 19590 | 19590 | 19590 |

## 4. Robustness checks

To assess the validity and sensitivity of our results, we implemented three sets of robustness checks. First, we dropped some key competition variables of our empirical model and analyzed the changes in the estimates of the remaining variables; second, we employed alternative estimators in addition to the 2SGMM utilized so far; and finally, similar to Greenfield (2014), we experimented with changing the concept of flight delays to inspect departure delays. The results are presented in Table 4 below and in the Appendix.

---

[22] Panel data of route-months for FSC carriers. Results produced by the two-step feasible efficient generalized method of moments estimator (2SGMM); statistics robust and efficient to arbitrary heteroscedasticity and autocorrelation; P-value representations: ***p<0.01, ** p<0.05, * p<0.10; results generated by alternative estimators presented in the Appendix.

Table 4 - Robustness checks[23]

|  | (1) ODDS | (2) ODDS | (3) ODDS | (4) ODDS | (5) ODDS | (6) ODDS | (7) ODDS |
|---|---|---|---|---|---|---|---|
| nr flights in congested hours | 0.0043* [0.002] | 0.0018 [0.002] | 0.0041* [0.002] | 0.0025 [0.002] | -0.0012 [0.003] | 0.0005 [0.001] |  |
| nr flights in uncongested hours | 0.0044** [0.002] | 0.0014 [0.001] | 0.0043** [0.002] | 0.0025* [0.001] | 0.0007 [0.003] |  |  |
| prop flights with bad weather | 4.7145*** [0.093] | 4.7425*** [0.090] | 4.7310*** [0.091] | 4.7321*** [0.089] |  | 4.7179*** [0.093] | 4.7174*** [0.093] |
| prop flights with incidents | 6.1807*** [0.395] | 6.1236*** [0.395] | 6.2785*** [0.397] | 6.3240*** [0.390] |  | 6.2095*** [0.397] | 6.2073*** [0.397] |
| prop flights held for late connections | 2.4781*** [0.258] | 2.3054*** [0.237] | 2.4142*** [0.256] | 2.3678*** [0.234] |  | 2.4464*** [0.253] | 2.4483*** [0.253] |
| max prop city delayed flights | 1.5460*** [0.229] | 1.6669*** [0.216] | 1.5449*** [0.220] | 1.6411*** [0.208] |  | 1.5606*** [0.225] | 1.5602*** [0.225] |
| codeshare agreement | 0.0081 [0.061] | 0.0173 [0.060] | 0.0099 [0.060] | 0.0094 [0.056] | -0.0508 [0.072] | 0.0054 [0.061] | 0.0047 [0.061] |
| HHI city-pair | 0.8192** [0.412] |  | 0.4247 [0.383] |  | 0.6137 [0.528] | 0.6491* [0.350] | 0.6535* [0.350] |
| HHI max endpoint cities | -1.5144*** [0.529] | -1.1549** [0.482] |  |  | -1.1624* [0.683] | -1.4143*** [0.527] | -1.4195*** [0.528] |
| LCC presence city-pair | -0.0412 [0.072] | -0.1616*** [0.037] | -0.0519 [0.070] | -0.1171*** [0.031] | 0.0155 [0.095] | -0.0669 [0.062] | -0.0660 [0.061] |
| LCC presence max endpoint cities | -0.4234** [0.180] | -0.4571** [0.179] | -0.2296* [0.135] | -0.1664 [0.134] | -0.4942** [0.216] | -0.4257** [0.180] | -0.4259** [0.180] |
| city-pair fixed effects | yes | yes | yes | yes | yes | yes | yes |
| time fixed effects | yes | yes | yes | yes | yes | yes | yes |
| Adj. R-Squared | 0.6800 | 0.6888 | 0.6863 | 0.6902 | 0.5036 | 0.6827 | 0.6826 |
| RMSE Statistic | 0.5890 | 0.5808 | 0.5831 | 0.5787 | 0.7335 | 0.5865 | 0.5865 |
| F Statistic | 76.576 | 78.720 | 78.241 | 79.564 | 44.064 | 77.509 | 77.705 |
| KP Statistic | 139.2305 | 320.8510 | 154.0443 |  | 136.5746 | 186.7474 | 186.8591 |
| KP P-Value | 0.0001 | 0.0001 | 0.0001 |  | 0.0001 | 0.0001 | 0.0001 |
| J Statistic | 3.2199 | 7.6572 | 11.9420 |  | 12.9733 | 3.0992 | 3.1458 |
| J P-Value | 0.3589 | 0.1050 | 0.0178 |  | 0.0047 | 0.3766 | 0.3697 |
| Weak CD Statistic | 83.4722 | 346.6552 | 85.5286 |  | 83.5622 | 112.2317 | 112.2612 |
| Weak KP Statistic | 30.5968 | 76.1872 | 33.8077 |  | 29.9950 | 40.6805 | 40.7102 |
| Nr Observations | 19419 | 19419 | 19419 | 19590 | 19419 | 19419 | 19419 |

The results in Table 4 show that our baseline model of $ODDS$ is robust to most changes in specifications. In particular, regarding the hypothesis of self-internalization, the results are notably insensitive to the omission of the competition variables. However, a comparison of the results among specifications (1) and (3) shows that $HHI\ city\text{-}pair_{kt}$ and $HHI\ max\ endpoint\ cities_{kt}$ are positively correlated and that omission of the second damages the proper estimation of the first[24]. This is indicative of a negative bias when omitting $HHI\,max\,endpoint\,cities_{kt}$. As the previous literature has not completely specified the empirical models regarding both these local and global indicators of concentration, we conclude that any subspecification related to concentration variables may cause relevant problems of inconsistent estimation and inference towards an unpleasant false negative. The same issue arises when comparing specification (1) and (5), and with $LCC\,presence\,max\,endpoint\,cities_{kt}$ in specification (4).

---

[23] Panel data of route-month for FSC carriers. Results produced by the two-step feasible efficient generalized method of moments estimator (2SGMM); statistics robust and efficient to arbitrary heteroskedasticity and autocorrelation; P-value representations: ***p<0.01, ** p<0.05, * p<0.10; results generated by alternative estimators presented in the Appendix.

The main results are not changed when we perform the additional robustness checks presented in the Appendix. In particular, the results do not change either when we run an LIML estimator or when we use departure delays instead of arrival delays in the regressand. The results are altered considerably, however, when employing the OLS estimator, as the estimated signs of $HHI$ change. We therefore recommend employing instrumental variables estimation to address endogeneity and its associated problems, as noted by Greenfield (2014).

**Conclusion**

The present paper estimated both the local and global effects of competition on the on-time performance of incumbent full-service carriers (FSCs) in the Brazilian airline market. By developing econometric models of the odds and the length of flight delays, we tested important relationships found in the recent literature as the hypotheses of airport congestion self-internalization and the market competition-service quality association. We also estimated the impacts of actual and potential competition with low cost carriers (LCC) in the industry.

Our results suggested that self-internalization of airport congestion was observed in the Brazilian airline market in the sample period. Additionally, it was also induced by LCC entry, with the effect of potential competition causing a positive spillover effect that benefitted non-entered routes towards better on-time performance. However, we only find that LCC entry benefits consumers by lowering the prevalence of flight delays, with no evidence found of any impact on the duration of such delays. We also find enough evidence supporting the competition-quality hypothesis in which lower concentration at the route level would enhance airline quality and thus forcing flight delays to decline.

---

[24] The Pearson correlation coefficient between the two variables was 0.48. See Table 2.

The interpretation of the combined results of the local and global effects of competitive conditions on the behavior of incumbent carriers regarding punctuality is as follows. In an apparent paradox, carriers tend to self-internalize congestion when their airport dominance is increased but also tend to maintain some self-internalization when this dominance is challenged by the entry of a LCC carrier. These movements are not contradictory as they may be an outcome of strategic scheduling adjustments provoked by the anticipation of vigorous price competition with a newcomer with a different business model. Our combined local and global effects therefore indicate that while quality competition regarding punctuality is observed locally in the market, there is evidence from our case study that the emergence and growth of LCCs may be an extra factor that enhances the on-time performance in the airline industry. However, there is certainly a need to consider applications of this model in other regions and also to consider further distinction among the different classes of LCCs around the world.

# Appendix

**Table 5 – Estimation results – LIML[25]**

|  | (1) ODDS | (2) ODDS | (3) MINS | (4) MINS | (5) MINS > 15 | (6) MINS > 15 |
|---|---|---|---|---|---|---|
| nr flights in congested hours | 0.0044** | 0.0044** | 0.0783* | 0.0801* | 0.0762* | 0.0786* |
|  | [0.002] | [0.002] | [0.041] | [0.044] | [0.042] | [0.044] |
| nr flights in uncongested hours | 0.0048*** | 0.0047** | 0.0451 | 0.0577 | 0.0452 | 0.0593 |
|  | [0.002] | [0.002] | [0.028] | [0.036] | [0.029] | [0.037] |
| prop flights with bad weather | 4.7164*** | 4.7187*** | 31.7594*** | 31.6361*** | 31.1087*** | 30.9740*** |
|  | [0.093] | [0.093] | [1.252] | [1.265] | [1.259] | [1.272] |
| prop flights with incidents | 6.2312*** | 6.1954*** | 50.0084*** | 50.5468*** | 48.4208*** | 48.9890*** |
|  | [0.396] | [0.398] | [5.023] | [5.038] | [5.136] | [5.147] |
| prop flights held for late connections | 2.4470*** | 2.4863*** | 22.2647*** | 20.1416*** | 22.1367*** | 19.8834*** |
|  | [0.277] | [0.259] | [5.114] | [4.530] | [5.170] | [4.573] |
| max prop city delayed flights | 1.5319*** | 1.5204*** | 12.4009*** | 12.3875*** | 12.2663*** | 12.2162*** |
|  | [0.226] | [0.229] | [2.916] | [2.932] | [2.948] | [2.968] |
| codeshare agreement | 0.0142 | 0.0024 | 1.1304 | 1.3488 | 1.1062 | 1.3325 |
|  | [0.062] | [0.061] | [0.904] | [0.871] | [0.908] | [0.876] |
| HHI city-pair | 0.8503** | 0.8699** | 30.4896*** | 30.0829*** | 31.8034*** | 31.5141*** |
|  | [0.381] | [0.419] | [8.722] | [9.045] | [8.874] | [9.220] |
| HHI max endpoint cities | -1.4278*** | -1.4609*** | -19.3882*** | -19.1038*** | -20.9400*** | -20.6756*** |
|  | [0.530] | [0.534] | [6.622] | [6.571] | [6.721] | [6.670] |
| LCC presence city-pair |  | -0.0280 |  | 2.4401 |  | 2.6603* |
|  |  | [0.073] |  | [1.498] |  | [1.529] |
| LCC presence max endpoint cities |  | -0.4092** |  | -0.5135 |  | -0.8489 |
|  |  | [0.180] |  | [1.694] |  | [1.718] |
| city-pair fixed effects | yes | yes | yes | yes | yes | yes |
| time fixed effects | yes | yes | yes | yes | yes | yes |
| Adj. R-Squared | 0.6795 | 0.6792 | 0.4557 | 0.4629 | 0.4387 | 0.4460 |
| RMSE Statistic | 0.5894 | 0.5897 | 6.1650 | 6.1243 | 6.2194 | 6.1792 |
| F Statistic | 78.859 | 76.443 | 26.479 | 26.753 | 25.189 | 25.448 |
| KP Statistic | 154.2698 | 139.2305 | 30.7876 | 29.8792 | 30.7876 | 29.8792 |
| KP P-Value | 0.0001 | 0.0001 | 0.0001 | 0.0001 | 0.0001 | 0.0001 |
| J Statistic | 3.1100 | 3.2160 | 0.2155 | 0.6474 | 0.1444 | 0.5478 |
| J P-Value | 0.3750 | 0.3595 | 0.6425 | 0.4210 | 0.7039 | 0.4592 |
| Weak CD Statistic | 91.0127 | 83.4722 | 35.3711 | 37.2128 | 35.3711 | 37.2128 |
| Weak KP Statistic | 34.0972 | 30.5968 | 10.3854 | 10.0724 | 10.3854 | 10.0724 |
| Nr Observations | 19419 | 19419 | 19590 | 19590 | 19590 | 19590 |

---

[25] Panel data of route-month for FSC carriers. Results produced by the limited-information maximum likelihood estimator (LIML); statistics robust and efficient to arbitrary heteroscedasticity and autocorrelation; P-value representations: ***p<0.01, ** p<0.05, * p<0.10.

**Table 6 – Estimation results – OLS[26]**

|  | (1) ODDS | (2) ODDS | (3) MINS | (4) MINS | (5) MINS > 15 | (6) MINS > 15 |
|---|---|---|---|---|---|---|
| nr flights in congested hours | 0.0022 [0.002] | 0.0016 [0.002] | 0.0109 [0.033] | 0.0064 [0.033] | 0.0054 [0.032] | 0.0010 [0.032] |
| nr flights in uncongested hours | 0.0027* [0.001] | 0.0014 [0.001] | -0.0225* [0.013] | -0.0334** [0.014] | -0.0257* [0.013] | -0.0365*** [0.014] |
| prop flights with bad weather | 4.7278*** [0.089] | 4.7385*** [0.089] | 32.1218*** [1.219] | 32.2104*** [1.215] | 31.4908*** [1.215] | 31.5801*** [1.211] |
| prop flights with incidents | 6.3678*** [0.392] | 6.3062*** [0.392] | 50.1286*** [4.495] | 49.6672*** [4.528] | 48.5654*** [4.544] | 48.0825*** [4.577] |
| prop flights held for late connections | 2.2006*** [0.239] | 2.3032*** [0.234] | 13.6712*** [3.805] | 14.5348*** [3.827] | 13.1073*** [3.813] | 13.9719*** [3.835] |
| max prop city delayed flights | 1.6748*** [0.207] | 1.6945*** [0.206] | 16.6318*** [2.241] | 16.8295*** [2.232] | 16.7074*** [2.237] | 16.8936*** [2.228] |
| codeshare agreement | 0.0292 [0.055] | 0.0132 [0.056] | 1.7922** [0.777] | 1.6754** [0.784] | 1.8006** [0.775] | 1.6769** [0.781] |
| HHI city-pair | -0.2086*** [0.064] | -0.3126*** [0.068] | 3.3899*** [0.627] | 2.4833*** [0.696] | 3.2958*** [0.630] | 2.3998*** [0.698] |
| HHI max endpoint cities | 0.1614 [0.201] | 0.1057 [0.196] | -2.4930* [1.384] | -2.9092** [1.378] | -2.9449** [1.391] | -3.3809** [1.384] |
| LCC presence city-pair |  | -0.1636*** [0.033] |  | -1.4236*** [0.351] |  | -1.4081*** [0.350] |
| LCC presence max endpoint cities |  | -0.1793 [0.137] |  | -0.6575 [1.340] |  | -0.9704 [1.343] |
| city-pair fixed effects | *yes* | *yes* | *yes* | *yes* | *yes* | *yes* |
| time fixed effects | *yes* | *yes* | *yes* | *yes* | *yes* | *yes* |
| Adj. R-Squared | 0.6900 | 0.6911 | 0.5599 | 0.5611 | 0.5554 | 0.5566 |
| RMSE Statistic | 0.5789 | 0.5779 | 5.5438 | 5.5362 | 5.5353 | 5.5278 |
| F Statistic | 79.059 | 79.180 | 32.814 | 33.438 | 32.745 | 33.437 |
| Nr Observations | 19590 | 19590 | 19590 | 19590 | 19590 | 19590 |

---

[26] Panel data of route-month for FSC carriers. Results produced by the Ordinary Least Squares estimator; statistics robust and efficient to arbitrary heteroscedasticity and autocorrelation; P-value representations: ***p<0.01, ** p<0.05, * p<0.10.

**Table 7 – Estimation results – departures[27]**

|  | (1) ODDSD | (2) ODDSD | (3) MINSD | (4) MINSD | (5) MINSD > 15 | (6) MINSD > 15 |
|---|---|---|---|---|---|---|
| nr flights in congested hours | 0.0043* | 0.0040* | 0.0826** | 0.0888** | 0.0813** | 0.0881** |
|  | [0.002] | [0.002] | [0.040] | [0.042] | [0.040] | [0.043] |
| nr flights in uncongested hours | 0.0045** | 0.0037* | 0.0600** | 0.0755** | 0.0602** | 0.0766** |
|  | [0.002] | [0.002] | [0.027] | [0.035] | [0.027] | [0.035] |
| prop flights with bad weather | 4.3423*** | 4.3484*** | 29.7430*** | 29.6021*** | 28.7533*** | 28.6096*** |
|  | [0.095] | [0.094] | [1.254] | [1.270] | [1.264] | [1.280] |
| prop flights with incidents | 5.6676*** | 5.6213*** | 45.6334*** | 46.2999*** | 42.9812*** | 43.6799*** |
|  | [0.394] | [0.393] | [4.931] | [5.006] | [4.976] | [5.055] |
| prop flights held for late connections | 2.3209*** | 2.4174*** | 23.4195*** | 21.5465*** | 23.3561*** | 21.4804*** |
|  | [0.271] | [0.251] | [5.145] | [4.646] | [5.176] | [4.672] |
| max prop city delayed flights | 1.4907*** | 1.4948*** | 11.9374*** | 11.8084*** | 11.6938*** | 11.5244*** |
|  | [0.227] | [0.228] | [2.818] | [2.847] | [2.827] | [2.861] |
| codeshare agreement | -0.0053 | -0.0198 | 1.0378 | 1.2075 | 0.9230 | 1.0905 |
|  | [0.065] | [0.065] | [0.891] | [0.868] | [0.892] | [0.870] |
| HHI city-pair | 0.5076 | 0.4762 | 28.9275*** | 29.7925*** | 29.3745*** | 30.4138*** |
|  | [0.372] | [0.409] | [8.107] | [8.538] | [8.185] | [8.643] |
| HHI max endpoint cities | -1.3852*** | -1.4100*** | -20.8782*** | -20.7291*** | -22.7749*** | -22.6519*** |
|  | [0.522] | [0.524] | [6.312] | [6.345] | [6.377] | [6.417] |
| LCC presence city-pair |  | -0.1175* |  | 2.3697* |  | 2.4545* |
|  |  | [0.071] |  | [1.413] |  | [1.431] |
| LCC presence max endpoint cities |  | -0.2771 |  | -0.0895 |  | -0.2276 |
|  |  | [0.183] |  | [1.618] |  | [1.634] |
| city-pair fixed effects | yes | yes | yes | yes | yes | yes |
| time fixed effects | yes | yes | yes | yes | yes | yes |
| Adj. R-Squared | 0.6793 | 0.6801 | 0.4566 | 0.4538 | 0.4438 | 0.4398 |
| RMSE Statistic | 0.5920 | 0.5913 | 6.0336 | 6.0497 | 6.0633 | 6.0858 |
| F Statistic | 85.521 | 83.803 | 25.775 | 25.397 | 24.419 | 23.988 |
| KP Statistic | 154.5998 | 139.3322 | 31.7590 | 30.4940 | 31.7590 | 30.4940 |
| KP P-Value | 0.0001 | 0.0001 | 0.0001 | 0.0001 | 0.0001 | 0.0001 |
| J Statistic | 2.6633 | 3.2374 | 0.1031 | 0.0008 | 0.1819 | 0.0167 |
| J P-Value | 0.4465 | 0.3564 | 0.7482 | 0.9769 | 0.6698 | 0.8973 |
| Weak CD Statistic | 91.3573 | 82.9942 | 36.8354 | 38.0770 | 36.8354 | 38.0770 |
| Weak KP Statistic | 34.2059 | 30.5977 | 10.7250 | 10.2840 | 10.7250 | 10.2840 |
| Nr Observations | 19408 | 19408 | 19579 | 19579 | 19579 | 19579 |

---

[27] Panel data of route-month for FSC carriers. Results produced by the two-step feasible efficient generalized method of moments estimator (2SGMM); statistics robust and efficient to arbitrary heteroscedasticity and autocorrelation; P-value representations: ***p<0.01, ** p<0.05, * p<0.10.